\documentclass[aps,pra,twocolumn,showpacs,groupedaddress,floatfix]{revtex4}
\usepackage{graphics}

\begin{document}

\title{Bose-Einstein Condensation on a Permanent-Magnet Atom Chip}
\author{C. D. J. Sinclair}
\author{E. A. Curtis}
\author{I. Llorente Garcia}
\author{J. A. Retter}
\altaffiliation{Present address: Laboratoire Charles Fabry de l'Institut d'Optique, UMR8501 du CNRS, 91403 Orsay Cedex, France.}%
\author{B. V. Hall}
\altaffiliation{Present address: Centre for Atom Optics and
Ultrafast Spectroscopy, Swinburne University of Technology,
Melbourne, Australia.}
\author{S. Eriksson}
\author{B. E. Sauer}
\author{E. A. Hinds}
\email{ed.hinds@imperial.ac.uk}

\affiliation{Centre for Cold Matter, Blackett Laboratory, Imperial
College, Prince Consort Road, London SW7 2BW, United Kingdom}

\begin{abstract}
We have produced a Bose-Einstein condensate on a permanent-magnet
atom chip based on periodically magnetized videotape. We observe
the expansion and dynamics of the condensate in one of the
microscopic waveguides close to the surface. The lifetime for
atoms to remain trapped near this dielectric material is
significantly longer than above a metal surface of the same
thickness. These results illustrate the suitability of microscopic
permanent-magnet structures for quantum-coherent preparation and
manipulation of cold atoms.
\end{abstract}


\maketitle

Atom chips are making rapid progress in the quantum-coherent
manipulation of microscopic cold atom clouds
~\cite{Folman,ReichelReview,HindsReview} with a view to
fundamental studies of quantum gases~\cite{QGLD},
interferometry~\cite{WangAnderson}, and quantum information
processing~\cite{Calarco}. Small-scale magnetic field patterns for
atom chips can be made using either microfabricated
current-carrying wires or microscopic structures of permanent
magnetization. This second idea is attractive because there is no
power dissipation in a permanent magnet and because very tight
atom traps with oscillation frequencies of $\sim 1\,$MHz (10\,nm
ground state size) are possible~\cite{Eriksson}. Previous studies
have used audiotape~\cite{Roach}, floppy
disks~\cite{Hughes1,Saba}, videotape~\cite{Rosenbusch1}, magnetic
and magneto-optical films~\cite{Spreeuw,Hannaford,Eriksson}, and
hard disks~\cite{Lev}. Until recently, permanently magnetized
films were used only for mm- or cm-scale manipulation of atom
clouds by reflection, but now cold atom clouds have been loaded
into the microtraps~\cite{Sinclair,Spreeuw}. For many applications
of these magnetic microtraps, the next significant step is to
prepare a Bose-Einstein condensate (BEC) on the chip as a source
of coherent matter waves for interferometry or as a low-entropy
reservoir for quantum information processing.

In this paper we describe the production of a BEC on a
permanent-magnet atom chip made from videotape, which forms an
array of waveguides. We have observed the propagation of the BEC
along one of these guides. With the ends of the guide closed, we
study center-of-mass and length oscillations of the trapped gas,
demonstrating that the videotape chip is a practical way to
manipulate cold atoms. We also show that the spin relaxation time
of the trapped atoms is significantly longer above the dielectric
surface of the videotape than above a metal surface of the same
thickness.

\begin{figure}
\resizebox{0.92\columnwidth}{!}{
\includegraphics{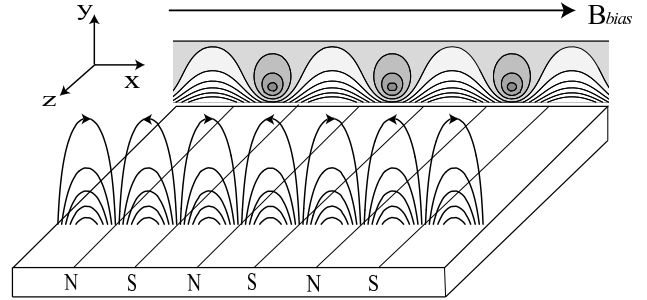}
    }
 \caption{Magnetic field lines produced by the
 magnetized videotape are shown in the foreground. A uniform bias field, $B_{bias}$, is added to this to make
 an array of atom guides. Contours of constant magnetic field
strength are shown in the background. Circular contours enclose
the lines of minimum field strength, where atoms are trapped.}
\label{fig:fieldlines}
\end{figure}

The waveguides of our atom chip are due to a sinusoidal pattern of
magnetization written along the length of the videotape ($\hat{x}$
direction) with the form $M_0\cos(kx)\hat{x}$, as illustrated in
Fig. \ref{fig:fieldlines}. This produces a field
\begin{equation}\label{equ:magfield}
(B_x,B_y) = B_{sur} e^{-ky}(-\cos(kx),\sin(kx)),
\end{equation}
where $B_{sur}$ is the field strength at the surface ($y=0)$. The
waveguides appear, as shown at the top of
Fig.~\ref{fig:fieldlines}, when a bias field $B_{bias}$ is added
in the $x$-$y$ plane. Near the center of each guide, the magnetic
field has a quadrupole structure with a gradient of field strength
given by $B^{\prime} = k B_{bias}$. An axial bias $B_z$ prevents
the total field from going to zero at the center. For
small-amplitude transverse oscillations this makes a harmonic trap
with frequency
\begin{equation}\label{equation:omega}
2 \pi f_{r} = k B_{bias}\sqrt{\frac{\mu_{B} g_F m_F}{m B_{z}}},
\end{equation}
where $\mu_{B} g_F m_F$ is the usual factor in the Zeeman energy
and $m$ is the mass of the atom.

Our atom chip is made using Ampex 398 Betacam SP videotape, which
has a 3.5$\,\mu$m-thick magnetic layer containing iron-composite
needles set in a glue and supported on a polymer ribbon 11\,$\mu$m
thick. The magnetization has a period of 106\,$\pm$\,2\,$\mu$m and
gives a surface field of $B_{sur}=11\pm1\,$mT\,\cite{Sinclair}.
With a small bias field of $B_{bias}=0.1$\,mT and $B_{z}=0.1$\,mT,
the atom guides lie $79\,\mu$m from the surface of the chip. For
ground state $^{87}$Rb atoms in the $|F=2,m_{F}=2\rangle$ sublevel
the transverse oscillation frequency is $760\,$Hz. The bias field
does not alter the magnetization of the videotape, which has very
low susceptibility and high coercivity. Further details on
permanent magnetic patterns may be found in
Ref.~\cite{HindsReview}.

\begin{figure}
 \resizebox{0.9\columnwidth}{!}{\includegraphics{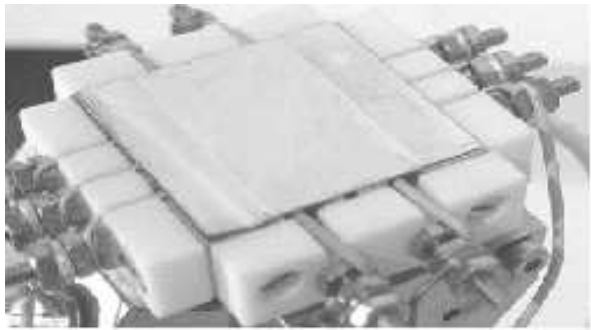}}
 \caption{Videotape atom chip assembly. The gold-coated surface is $2.5 \times 2.5$ cm.
 Wires underneath facilitate the loading and manipulation of atoms on the chip. }
 \label{atomchip}
 \end{figure}

A piece of recorded videotape is glued to a glass coverslip and is
coated with approximately $400$\,nm of gold to make the surface
reflect 780\,nm light. This allows the formation of a
magneto-optical trap (MOT) by reflection\,\cite{ReichelHaensch} in
order to collect and cool the atoms close to the videotape. The
coverslip is then glued to a stainless-steel base to form the
assembly shown in Fig.~\ref{atomchip}. Underneath this, three
wires run parallel to the $z$ direction. The central one, the
``center wire'', is used to transport cold atoms from the MOT into
one of the videotape guides, whilst the outer two form an rf
antenna for evaporative cooling. The two wires running along $x$
allow the guide to be pinched off to form a trap by making a field
whose $z$ component rises to a maximum above each wire. The whole
assembly is mounted in a vacuum chamber, where a remarkably low
outgassing rate~\cite{Hopkins} allows the vacuum to reach
$\sim10^{-11}\,$Torr. Further details of the construction are
given in Ref.\,\cite{Sinclair}.

Cold $^{87}$Rb atoms from a low-velocity intense source
(LVIS)\cite{Lu} are captured 4\,mm from the surface by the
reflection MOT. This is loaded for $20\,$s to collect typically
$10^9$ atoms before the LVIS is turned off. The MOT is moved to
1.5\,mm from the surface by ramping up an external bias field,
while increasing the  laser detuning from -15\,MHz to -45\,MHz,
cooling the cloud to $50\,\mu$K. Next, the MOT light and
quadrupole field are switched off, the atoms are optically pumped
into the $|F=2,m_F=2\rangle$
 state, and the cloud is recaptured in a purely magnetic trap, formed by
passing 15\,A through the center wire with a 1.4\,mT bias field
along $x$.

The bias field is ramped up to 4.4 mT over 100 ms, which increases
the trap depth, compresses the cloud and moves it to 160\,$\mu$m
from the surface. An initial stage of forced evaporative cooling
ensures the cloud is cold enough to load only one of the videotape
microtraps. The rf field is swept over 6\,s from 30\,MHz to
3.9\,MHz, cooling the cloud to $\sim 15 \mu$K. We continue to
evaporate at 3.9\,MHz by reducing the center-wire current and the
bias field over 4\,s. This gradually merges the wire trap with one
of the videotape microtraps as the cloud approaches the surface.
The wire current is then ramped to zero leaving the atoms confined
entirely by the videotape microtrap at a temperature of order
$10\,\mu$K.

 \begin{figure}
 \resizebox{1\columnwidth}{!}{\includegraphics{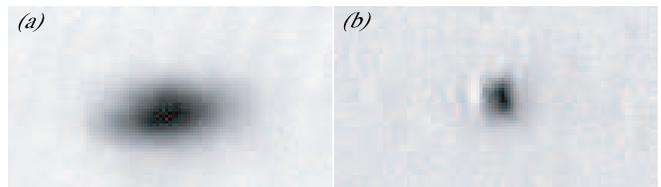}}
 \caption{Absorption images taken after 8.7 \,ms of free expansion.
 (a) Thermal cloud at 500 nK with a final rf frequency of 820 kHz.
 (b) Condensate with a final rf frequency of 810 kHz.
 Image area is $0.9\times 0.5$ mm.}
 \label{BECfree}
 \end{figure}

A final rf sweep lasting two or three seconds, typically down to
$800\,$kHz, cools the cloud below its critical temperature. The
transition to a condensate can be seen by releasing the cloud,
allowing it to expand, and imaging its density by the absorption
of light tuned to the D2($F=2\rightarrow 3$) transition.
Figure\,\ref{BECfree}(a) shows a 500\,nK thermal cloud imaged
after 8.7\,ms of expansion. The trap was $88\,\mu$m from the
surface with $f_r =450\,$Hz and $f_z =15\,$Hz giving a prolate
aspect ratio of $30:1$. The isotropic thermal velocity
distribution gives a nearly spherical image after expansion. By
contrast, the condensate imaged in Fig.~\ref{BECfree}(b) is
oblate, as expected for a condensate that is prolate in the
microtrap. The cloud is released by switching off $B_{bias}$ in a
time of 3 to 4 ms.

\begin{figure}
 \resizebox{0.87\columnwidth}{!}{\includegraphics{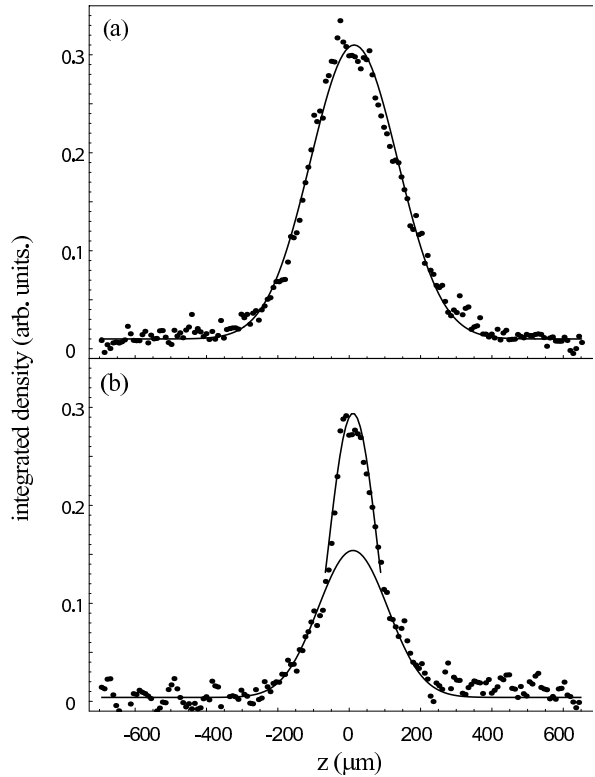}}
 \caption{Atom density profiles measured by absorption imaging after
allowing the cloud to expand for 13.7 ms in a videotape waveguide.
(a) Thermal cloud at 520\,nK. (b) Partially condensed cloud with a
thermal component at 330\,nK. The parabolic central peak
corresponding to the condensed fraction is $180\,\mu$m wide.}
 \label{BECaxial}
 \end{figure}

An important feature of the atom chip is that it creates
waveguides in which the atoms can propagate. We have used this as
a better way to see the formation of the condensate. After
evaporation, we open the ends of the trap by switching off the end
wires. Now the cloud expands freely along the guide and we make
absorption images of the atoms in situ. Figure~\ref{BECaxial}(a)
shows the Gaussian density profile of a 520\,nK thermal cloud
after 13.7\,ms of expansion in a guide with $f_r=580\,$Hz. By
contrast, Fig.~\ref{BECaxial}(b) shows a similarly expanded BEC.
Here the thermal component has a temperature of 330\,nK as
obtained from a fit to the wings of the profile. Near the center
we add a parabola of $180\,\mu$m length to this in order to fit
the region of the condensate. The expected condensate length can
be found by the following rough estimate. Since a quarter of the
cloud is condensed at 330\,nK, we calculate that the initial trap
($f_r =320\,$Hz, $f_z =15\,$Hz) has about $8\times 10^4$ atoms in
the condensate, with a Thomas-Fermi length of $90\,\mu$m. After
13.7\,ms of expansion (see Eq. 11 of Ref.~\cite{CastinDum}) this
length is expected to grow to $200\,\mu$m. Our measurement
therefore indicates that the interaction with the thermal cloud
may slightly impede the expansion, although we do not currently
have enough accuracy in all the relevant parameters to be sure. In
any case, the condensate length grows rapidly in the waveguide
compared with a 3D expansion because there is no radial expansion
to release the mean-field energy.

\begin{figure}
 \resizebox{1\columnwidth}{!}{\includegraphics{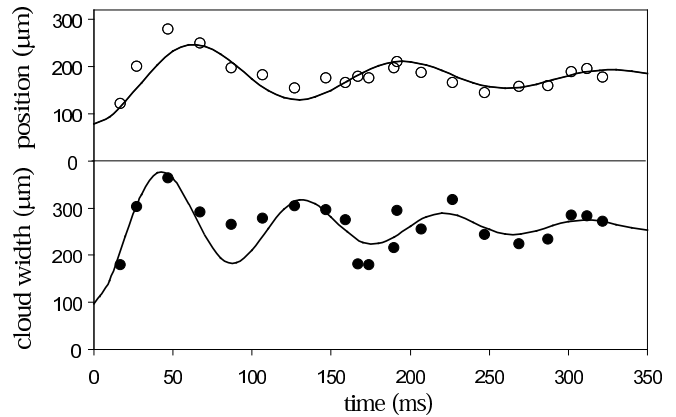}}
 \caption{Oscillations following a jump of
the axial frequency. Open circles: center-of-mass oscillation.
Filled circles: length oscillation. Solid curves are damped
cosines. Each point represents a single destructive measurement.}
\label{oscillation}
 \end{figure}

The permanent-magnet waveguides also offer a suitable setting in
which to study more complex dynamics of the cloud. In order to
demonstrate this possibility we jump the axial frequency to a
lower value but do not remove the axial trapping entirely. This
excites a length oscillation of the trapped
cloud~\cite{JinCornell,ZimmermannOsc}. We also see an oscillation
of its center-of-mass, due to a gravitational shift of the
equilibrium position because the atom chip is not perfectly
horizontal. Open circles in Fig.~\ref{oscillation} show the
center-of-mass motion, which is well described by a damped cosine
of frequency $f_z=7.6\pm0.4\,$Hz and $\sim 180\,$ms decay time.
The damping is probably due to a slight roughness in the magnetic
potential, which results from defects in the videotape
material~\cite{Sinclair}. When the initial cloud is thermal, we
measure a frequency for length oscillations of $2f_z$, as
expected. By contrast, if we prepare a partly condensed cloud
before relaxing the trap we measure a slower breathing frequency.
After a few tens of milliseconds, the atom density profiles no
longer show a clear two-component distribution, indicating
coupling between the condensed and thermal components. The filled
circles in Fig.~\ref{oscillation} show how the length of the cloud
evolves over 300\,ms. The main features of the motion are once
again described by a damped cosine with $180\,$ms decay time, but
now the frequency is $11.2\pm0.4\,$Hz. Within the accuracy of our
measurements this frequency is consistent with $\sqrt{12/5}f_z =
11.8\pm0.6\,$Hz as expected for a Bose gas in the hydrodynamic
limit~\cite{GriffinStringari,StamperKurnKetterle}.  We anticipate
that the damping time of the length oscillation would be longer in
a smoother guide and with a pure condensate. We are now studying
the effects of roughness on the dynamics of cold clouds in these
waveguides and we are pressing towards guides of higher radial
frequency where aspects of 1D many-body physics may be explored.

\begin{figure}
 \resizebox{0.9
 \columnwidth}{!}{\includegraphics{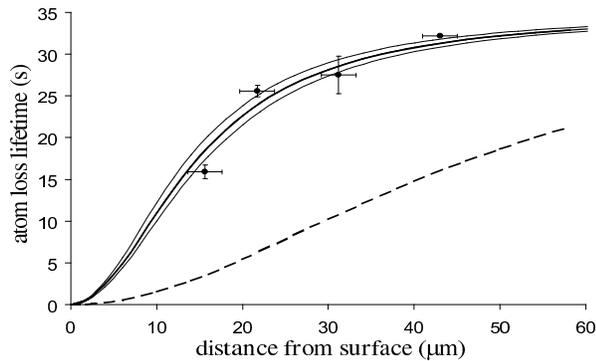}}
 \caption{Data: lifetimes for loss of atoms trapped above videotape atom chip with
  a $400\,$nm gold layer. Solid lines: theory of Scheel \textit{et.
  al.} at temperature $350\pm 50$K. Dashed line: expected lifetime near a $4\,\mu$m-thick gold layer.}
 \label{lifetime}
 \end{figure}

In contrast to other atom chips, the material of the videotape is
an insulator, not a conductor. This is significant because the
trapped atom loss due to spin flips is expected to be much less
severe near an insulating surface~\cite{Harber2003} than it is
near a metal~\cite{Jones,LinVuletic}. Under conditions that apply
to our experiment, we expect a lifetime $\propto \rho d^2 /h$,
where $\rho$ is the resistivity of the film, $h$ is its thickness,
and $d$ is the atom-surface distance~\cite{Folman}. At the MHz
frequencies of relevance here, the resistivity of the videotape
film is at least a million times larger than that of gold, so it
is predicted that the spin flip rate above our chip should be due
entirely to the 400\,nm gold film on the surface. Scheel
\textit{et al.} have recently derived a rigorous formula for this
rate~\cite{Rekdal,Scheel}. Figure~\ref{lifetime} shows the
lifetime calculated according to Ref.~\cite{Scheel} due only to
the 400\,nm gold layer at the $350(\pm50)\,$K temperature of the
chip. Far from the chip the lifetime is 35\,s due to collisions
with the background gas. The data points show the longest thermal
cloud lifetimes we have measured at various heights above the
chip. They agree, confirming that the videotape itself makes no
appreciable contribution to the loss rate. For comparison, the
dashed line shows the calculated lifetime near a $4\,\mu$m-thick
gold layer, which is the total thickness of our magnetic layer
plus the gold coating, and is also typical of the gold thickness
used to make atom chips with current-carrying wires. This result
shows that permanent magnets may be preferable to metal atom chips
in applications where the atoms' quantum states decohere as a
result of magnetic field fluctuations.

In this paper we have shown that a long thin BEC can be prepared
in a microtrap formed by a videotape atom chip and can be
manipulated in the waveguides of the chip. We have demonstrated a
long spin-relaxation lifetime for atoms trapped in this way. These
experiments show that permanent-magnet micro-structures offer a
powerful new method for making atom microtraps with long lifetimes
and tight confinement, which are relevant for applications
involving low-dimensional quantum gases. To build more complicated
quantum circuits, such as those required for quantum information
processing, it will be necessary (and straightforward
\cite{Eriksson}) to write more elaborate magnetic patterns on the
surface.

\begin{acknowledgments}
We acknowledge the technical assistance of Jon Dyne.  This work is
supported by the UK EPSRC Physics Program, the RCUK Basic
Technology Program, the Royal Society and the QGates, AtomChips,
and FASTnet networks of the European Commission.
\end{acknowledgments}

\end{document}